\documentclass[twocolumn,showpacs,amsmath,amssymb,pre,aps]{revtex4}   %%  4-pages !!
\usepackage{graphicx}% Include figure files
\usepackage{dcolumn}% Align table columns on decimal point
\usepackage{bm}% bold math

\begin{document}
\title{
Nonuniform Self-Organized Dynamical States in Superconductors with Periodic Pinning
}
\draft

\author{Vyacheslav R. Misko$^{1,2,3}$, Sergey Savel'ev$^{1,4}$, Alexander L. Rakhmanov$^{1,5}$, 
and Franco Nori$^{1,2}$}
\affiliation{$^1$ Frontier Research System, The Institute of Physical and Chemical Research (RIKEN), Wako-shi, Saitama, 351-0198, Japan}
\affiliation{$^2$ MCTP, Department of Physics, University of Michigan, Ann Arbor, MI 48109-1040, USA}
\affiliation{$^3$ Department of Physics, University of Antwerpen (CGB), B-2020 Antwerpen, Belgium}
\affiliation{$^4$ Department of Physics, Loughborough University, Loughborough LE11 3TU, United Kingdom}
\affiliation{$^5$ Institute for Theoretical and Applied Electrodynamics Russian Academy of Sciences, 125412 Moscow, Russia}

\date{\today}

\bigskip

\begin{abstract}
We consider magnetic flux moving in superconductors with periodic
pinning arrays. We show that sample heating by moving vortices
produces negative differential resistivity (NDR) of {\it both N
and S} type (i.e., {\it N}- and {\it S}-shaped) in the voltage-current 
characteristic
(\textit{VI} curve). The uniform flux flow state is unstable in
the NDR region of the \textit{VI} curve. Domain structures appear
during the NDR part of the \textit{VI} curve of an $N$ type, while a
filamentary instability is observed for the NDR of an $S$ type. The
simultaneous existence of the NDR of both types gives rise to the
appearance of striking self-organized (both stationary and
non-stationary) two-dimensional dynamical structures.
\end{abstract}


\maketitle

Semiconductor devices~\cite{sch} exhibiting Negative Differential
Resistivity (NDR) and Conductivity (NDC) have played a very important
role in science and technology.
These useful devices include~\cite{sch}: Gunn effect diodes, $pnpn$-junctions, etc.
Here we study superconducting analogs of these semiconductor devices.
Table~1 briefly compares NDR in superconductors, semiconductors, plasmas, 
and manganites.
Conceptually, the non-uniform self-organized structures
(e.g., filaments and overheated domains with higher or lower electric
fields) are related for superconductors, plasmas, and semiconductors.
However, the physical mechanism giving rise to the instability of the
homogeneous state can be different in each case.

The magnetic flux behavior in superconductors with artificial
pinning sites has attracted considerable attention due to
the possibility of constructing samples with desired properties
\cite{vvmdotprl,fnsc2003,rwdot,vvmbdot,vvmfddot}.
Among such systems, samples with
Periodic Arrays of Pinning Sites (PAPS) are studied intensely
because advanced
fabrication techniques allow to design well-defined periodic
structures with controlled microscopic pinning parameters.
For such systems,
Ref.~\onlinecite{fn1}
has revealed the existence of several dynamical vortex phases,
which are similar to the ones shown in the inset of Fig.~1.
At low current density $j$, the vortices are pinned and
their average velocity, $\bar{v}$, is zero (phase I).
Here
$\bar{v} = N_{v}^{-1} \sum_{i} \mathbf{v}_{i} \cdot \mathbf{x}$,
is the average vortex velocity in the $x$ direction.
At higher $j,$ interstitial vortices start to move and the flux velocity
increases with the drive (phase II). Then, a sharp jump in
$\bar{v}(j)$ occurs since a fraction of the vortices depins and a
very disordered uniform phase
arises (phase III).
This phase is analogous to the {\it uniform} electron
{\it motion in semiconductor devices}.
After increasing the applied current $j$ (for superconductors)
or the applied voltage (for semiconductors),
the vortex (electron) velocity shows a remarkable and
non-intuitive sudden drop
(i.e., the velocity {\it drops} even though the applied force {\it increases}).
Indeed, when $j$ exceeds a threshold value, only incommensurate
vortex rows move (phase IV). Finally, the increasing driving force
completely overcomes the pinning (phase V). The dependence
$\bar{v}(j)$ describes the voltage-current characteristic
(\textit{VI} curve) since the electric field $E$ is related to
$\bar{v}$ by $E=-\bar{v}(j)B/c$.
Here we prove that the flux motion in
samples with PAPS results in an unusual,
for superconductors,
\textit{VI} curve with
NDR of the so-called $S$ type \cite{sch,kun}.
That is, within some interval of
voltages there exist {\it three} different current values corresponding
to a {\it single} electric field.
Such types of $S$-shaped \textit{VI} curves play a very important role in
plasmas and semiconductors, and
give rise to a {\it filamentary instability} when a uniform current
flow breaks into filaments with lower and higher current
densities~\cite{sch}.

%\vspace{0.5cm}
\begin{table*}
\begin{tabular}{|c|c|c|c|c|} 
  \hline
  % after \\: \hline or \cline{col1-col2} \cline{col3-col4} ...
   & {\bf Superconductors} & {\bf Semiconductors} & {\bf Plasmas} & {\bf Manganites} \\ 
  \hline
  {\bf Carriers} & flux quanta & charge quanta:     & electrons & electrons, \\ 
                 &             & electrons or holes &           & holes      \\ 
\hline
  {\bf Characteristic curve} & voltage-current     & current-voltage  & {\it IV} curve & {\it IV} curve \\ 
                             & (\textit{VI}) curve & ({\it IV}) curve &                &                \\ 
\hline
  {\bf Homogeneous state} & homogeneous flux  & homogeneous  & homogeneous  & homogeneous  \\ 
                          & and current flows & current flow & current flow & current flow \\ 
\hline
  {\bf Origin of S-shape} & in/commensurate vortex   & non-linear electron & ionization & $-$ \\ 
  {\bf NDR}               & dynamical phases in PAPS & transport           &            &     \\ 
   \hline
  {\bf Origin of N-shape} & overheating, Cooper pair                                 & overheating, electron & heating \cite{cap} & heating \cite{tokunaga2} \\ 
  {\bf NDR}               & tunnelling~\cite{Hueb}; vortex-core:                     & or hole tunnelling    &         &         \\ 
                          & shrinkage~\cite{LO} ($T \approx T_{c}$);                 &                       &         &         \\ 
                          & expansion~\cite{kun}, driven~\cite{doettinger} ($T \ll T_{c}$) &                       &         &         \\ 
\hline
  {\bf Filaments} & supercurrent filaments & normal current filaments & pinch-effect & $-$ \\
\hline
  {\bf Domains} & vortex-induced higher $E$ & higher $E$            & higher $E$            & higher $E$            \\ 
                & field overheated domains  & overheated domains    & overheated domains    & overheated domains    \\ 
  \hline
\end{tabular}

%\vspace{0.3cm}
\caption{\label{tab:table1}
Comparison between non-uniform non-equilibrium states in superconductors, semiconductors, 
plasmas, and manganites with \textit{VI} curves having Negative Differential Resistivity (NDR).
Since $IV$ curves in semiconductors map to $VI$ curves in superconductors,
then NDC (for semiconductors) maps into NDR for superconductors.
Here, $N$($S$) type shapes for semiconductors correspond to $S$($N$) type for
superconductors~\cite{GMR}.
The Negative Differential Conductivity (NDC) found in~\cite{kun} is analogous to
the Gunn effect in semiconductors, where electron-charge modulations lead
to steps in $j(E)$ in the NDC regime.
}

\vspace{-0.5cm}
\end{table*}
%\vspace{-0.5cm}

Sufficiently strong
disorder in the pinning array gives rise to the disappearance of
the described dynamical phase transitions
\cite{fn1}
and, consequently, the vanishing of the NDR for $S$-shaped 
\textit{VI} curves.
Higher thermal fluctuations also result in such an effect.
Experimentally,
the current density, $j$, at which the flux flow regime is observed in
superconductors, is usually high~\cite{BL,GMR}, and the Joule heat
should also affect the picture described above. An increase in
temperature $T$ results in a decrease of the pinning force.
Thus, the current density can decay with increasing electric
field, and a $N$-shaped \textit{VI} curve with a NDR of $N$ type (red dashed line
in Fig.~1) is commonly observed in superconductors for high current
density~\cite{GMR,GM}. The uniform state in samples
with NDR of $N$ type is also unstable~\cite{sch}; and a propagating resistive
state boundary or the formation of resistive domains can destroy the
uniform superconducting state~\cite{GMR,GM,Hueb}.
For certain
pinning parameters and cooling conditions, one can achieve a
situation where the NDR of {\it both} $N$ and $S$ type {\it simultaneously coexist}
in the \textit{VI} curve (Fig.~1). In this case, we predict
remarkable
flux flow instabilities.

Here, we study the effect of Joule heating and disorder
on the \textit{VI} curve of superconductors with PAPS.
Based on
analytical and
numerical analysis of the \textit{VI} curve, we
find the conditions under which the \textit{VI} curve has a NDR
region of either $S$ type or $N$ type, or both. We discuss the effect of
the shape of the NDR on the vortex and current flow.
We argue
that the coexistence of the NDR of {\it both}, $S$ and $N$ types, gives
rise to the macroscopically non-uniform self-organized
dynamical structures in the flux flow regime.

We numerically integrate the two-dimensional overdamped equations of motion
\cite{fn1,md01,md03Z}
for the flux lines driven, by the current $j$, in the $x$ direction over a square
PAPS with lattice constant $a$:
$\eta \mathbf{v}_i=\mathbf{F}_i^{vv}+\mathbf{F}_i^{vp}+\mathbf{F}_{i}^{T}+\mathbf{F}_d.$
Here
$\eta$
is the flux flow viscosity~\cite{BL},
$\mathbf{v}_i$ is the
velocity of $i$th vortex, $\mathbf{F}_i^{vv}$ is the force acting
on the $i$th vortex per unit length due to the interaction with
other vortices, $\mathbf{F}_i^{vp}$ is the $i$th vortex-pin
interaction, $F_{d} = j \phi_{0} / c$ is
the driving force, and $\phi_0$ is the flux quantum. The thermal
fluctuation contribution to the force, $\mathbf{F}_{i}^{T}$,
satisfies:
$\langle F_{i}^{T}(t) \rangle_{t} = 0$ and $\langle
F_{i}^{T}(t)F_{j}^{T}(t^{\prime}) \rangle_{t} = 2 \eta k_{B} T
\delta_{ij} \delta(t-t^{\prime})$.

The vortex-vortex interaction is modelled by $
\mathbf{F}_i^{vv}=\left( \phi_0^2 / 8\pi^2\lambda^3(T) \right)
\sum_{j=1}^{N_v}K_1 \left( |\mathbf{r}_i-\mathbf{r}_j| /
\lambda(T) \right) \mathbf{\widehat{r}}_{ij}, $
$K_1$ is the modified Bessel
function, the summation is performed over the positions $\mathbf{r}_j$ of $N_v$
vortices in the sample, and
$\mathbf{\widehat{r}}_{ij}=(\mathbf{r}_i-\mathbf{r}_j)/|\mathbf{r}_i-\mathbf{r}_j|$.
The temperature dependence of
the penetration depth
$\lambda$ is approximated as
$\lambda(T)=\lambda_0(1-T^2/T_c^2)^{-1/2}$. The Ginzburg-Landau
formula for $H_{c2}(T)$ is used and we assume that the ratio
$\kappa_{GL}=\lambda/\xi$ is temperature independent (here $\xi$
is the coherence length). The pinning is modelled as $N_p$
parabolic wells located at positions $\mathbf{r}_k^{(p)}$. The pinning
force per unit length is $ \mathbf{F}_i^{vp}= \left( F_p(T) / r_p
\right) \sum_{k=1}^{N_p}|\mathbf{r}_i-\mathbf{r}_k^{(p)}|\Theta\left(
r_p-|\mathbf{r}_i-\mathbf{r}_k^{(p)}| / \lambda_0
\right)\mathbf{\widehat{r}}_{ik}^{(p)}, $ where $r_p$ is the range of
the pinning potential, $\Theta$ is the Heaviside step function,
and
$\mathbf{\widehat{r}}_{ik}^{(p)}=(\mathbf{r}_i-\mathbf{r}_k^{(p)})/|\mathbf{r}_i-\mathbf{r}_k^{(p)}|$.
We estimate the maximum pinning force, $F_p=F_p(T)$, as
$H_c^2\xi^2/r_p$ and, thus, $F_p(T)=F_{p0}(1-T^2/T_c^2)$.

For brevity,
the simulations shown here are for $18\times12$~$\lambda_0^2$ periodic cells
and at magnetic fields near the first matching
field, $B_{\phi}=\phi_0/a^2$, where $N_p=N_v$.
First,
we start with a high-temperature vortex liquid.
Then, the temperature
is slowly decreased down to $T = 0$. When cooling down, vortices adjust
themselves to minimize their energy, simulating field-cooled
experiments. Then we increase the driving current and compute the
average vortex velocity $\bar{v}(j)$.

\begin{figure}[btp]
\begin{center}
\vspace*{-7.0cm}
\hspace*{-1.0cm}
\includegraphics*[width=10.0cm]{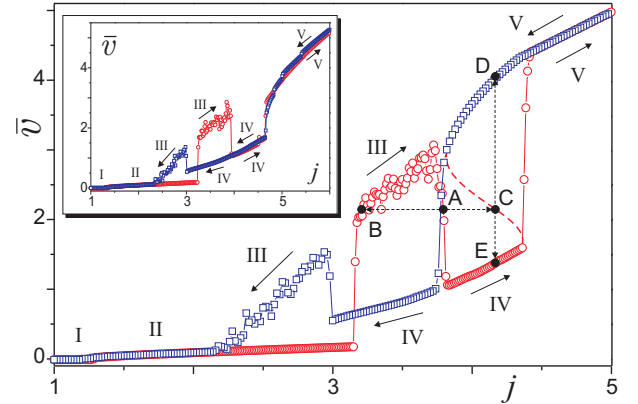}%[bb=94 490 434 704, width=8.0cm, clip]
\end{center}
\vspace{-2.0cm}
\caption{ (Color online)
The average vortex velocity
$\bar{v} \propto E$ vs current $j$ for
$B/B_{\Phi} = 1.074$, $r_{p} = 0.2\lambda_{0}$ and $F_{p}/F_{0} =
2$ for increasing (red open circles) and for decreasing (blue open
squares) $j$. State A is unstable and the sample divides
into filaments, some in states B and some in C. State C is also unstable.
The corresponding stable states are on the lower (point E) and on
the upper (point D) \textit{VI} curve branches.
Inset: {\it VI} curve when no heating effects are taken into account.}
\vspace{-0.5cm}
\end{figure}

The average power of Joule heating per unit volume is
$jE=j\bar{v}B/c$. We assume that the temperature relaxation length
is larger than any local scale and the sample thickness.
Under such conditions, the local temperature
increase due to vortex motion can be found
from the heat balance equation~\cite{GMR}:
$h_0S\left(T-T_0\right)=\bar{v}jBV/c$, where $h_0$ is the heat
transfer coefficient, $T_0$ the ambient temperature,
and $S$ and $V$ are the sample surface and
volume. Further, we shall assume
that $T_0\ll T_c$ and neglect $T_{0}$. We define the
dimensionless driving force as $f_d=j/j_0$ and introduce the
dimensionless parameters $V_x=\bar{v}/v_0$ and $b=B/B_{\phi}$,
where $v_0=c^2/4\pi\kappa^2_{GL}\sigma_n\lambda_0$, and
$j_0=c\phi_0/8\pi^2\lambda_0^3$. We assume that
the normal state conductivity $\sigma_n$ is temperature-independent.
As a result, the temperature and
the driving force are related by: $T/T_c=K_{th} \, V_x \, f_d \, b$, where
$K_{th}=j_0 \, v_0 \, B_{\phi} \, V/c \, h_0 \, T_c \, S$ is the ratio of
the characteristic heat release to heat removal.
Using the rough
estimates
$\lambda_0=2000$~\AA, $\kappa_{GL}=100$,
$\sigma_n=10^{16}$~s$^{-1}$, $V/S=1000$~\AA, $T_c=90$~K,
$B_{\phi}=500$~G, and $h_0=1$~W/cm$^2$K, we find
$K_{th}=0.05-0.06$ and $F_{p0}$ is of the order of
$F_0=\phi_0j_0/c$. In the simulations, we used $K_{th}=0.0525$.

The velocity $\bar{v}$ versus current $j$
(which coincides with the sample \textit{VI} curve in
dimensionless variables) is presented in Fig.~1 for increasing and
for decreasing $j$ (if we neglect the heating effect, $\bar{v}(j)$ has the shape shown
in the inset of Fig.~1, similar to Ref.~\cite{fn1}).
For low currents,
$j \lesssim 3$,
the effect of heating is negligible; for
$j \gtrsim 3.7$,
the behavior of $\bar{v}(j)$ drastically changes,
compared to the non-heating case shown in the inset of Fig.~1.
In particular,
an abrupt transition occurs between regimes IV and V.
The most pronounced
feature related to {\it heating} is the appearance of {\it hysteresis}
in regions IV and V: the overheated vortex lattice
(for decreasing $j$)
keeps moving
as a whole at lower currents than the ``cold'' one
(for increasing $j$).
As a result, we obtain a complex $N$ and $S$ type
\textit{VI} curve, characterized by {\it two kinds of
instabilities}. For example, if the current density exceeds the
value $j\approx 3.7$ (point A), the uniform current flow is
unstable and the so-called filamentary instability~\cite{sch}
occurs.
Consequently, the current flow breaks into
supercurrent
filaments, some with lower $j_B$ (state B) and others with higher $j_C$ (state C).
The state C is, in its turn, unstable. The
corresponding stable states are on the lower (E) and on the upper
(D) \textit{VI} curve branches.

Note that a small amount of pinning disorder influences $\bar{v}(j)$ and can 
lead to the disappearance of phase III \cite{wendrl}. 
However, {\it the robust hysteresis, related to heating, remains}.

For a sample included in an electric circuit (see inset in Fig.~2),
the circuit equation is $L\dot{I}+RI+lE=U$, where $I=jA$,
$L$ and $R$ are the inductance and resistance, $A$ and $l$ are the
sample cross-section and length, and $U$ is a constant voltage.
The circuit and Maxwell equations describe
the development of small field perturbations $\delta \mathbf{E}$
and $\delta \mathbf{B}$ and current $\delta j$.
We seek perturbations of the form $\delta \! E,\,\delta \! B,\,
\delta j \propto \exp(\nu t/t_0)$,
where $\nu$ is the value to be found and $t_0=L/R$.
An instability develops if $Re(\nu)> 0$.
In general, we
should consider also the thermal equation, but for filamentary
instabilities the temperature rise is not of importance. To find the
instability criterion we consider $y$-dependent perturbations~\cite{sch}.
In such a geometry, $\delta
\mathbf{E}$ has only the $x$ component, while $\delta \mathbf{B}$
has only the $z$ component.
Using $\delta \! B = c t_0 \delta E'/\nu w$,
we find from the circuit and Maxwell equations that:
\begin{eqnarray}
\label{A3} (\nu+1) \delta \bar{j}+\rho_c^{-1}\delta \bar{E} &=& 0, \\
\label{A4}  \delta E''-\beta \delta E'-\frac{\nu t_s}{t_0} \delta E &=& 0,
\end{eqnarray}
where $\delta \bar{j}$ and $\delta \! \bar{E}$ are the values
averaged over the cross-section, prime is differentiation over the
dimensionless coordinate $y/w$, $w$ is the sample half-width,
$\rho_c=RA/l$,
$ t_s=(4\pi w^2/c^2)(\partial j / \partial E) $,
$ \beta=(4\pi w / c)(\partial j / \partial B) $
are parameters, which are either positive or negative depending on
the part of the \textit{VI} curve for given background fields $E$ and $B$.
All quantities are averaged over a volume which includes a large number of vortices.

The first boundary condition to Eq.~(\ref{A4}), $\delta
B(w)=-\delta B(-w)$, is obtained assuming that the applied magnetic
field $B$ is constant. From Maxwell equations we derive $\delta
B(w)-\delta B(-w)=8\pi w \delta\bar{j}/c$.
Using these $\delta B$'s, and substituting $\delta \bar{j}$
from Eq.~(\ref{A3}), we obtain the second boundary condition
$\delta E'(w)=-\gamma\nu\delta \bar{E}/(\nu+1)$, where
$\gamma=4\pi w^2l/c^2AL$.
Substituting the solution of
Eq.~(\ref{A4}), 
$\delta E=C_1\exp(p_1y/w)+C_2\exp(p_2y/w),$
to the boundary conditions and requiring a zero
determinant for the obtained set of linear equations for
constants $C_i$, we find the equation for the
eigenvalues $\nu$
\begin{equation}
\label{A9}
\left[p_2+\frac{\gamma\nu}{(\nu+1)p_2}\right]\frac{p_1\tanh
p_2}{p_2\tanh p_1}=p_1+\frac{\gamma\nu}{(\nu+1)p_1} \ ,
\end{equation}
where $p_{1,2}=\beta/2\pm \sqrt{\beta^2/4+\nu t_s/t_0}$.

\begin{figure}[btp]
\begin{center}
\vspace*{-7.5cm}
\hspace*{-1.0cm}
\includegraphics*[width=10.0cm]{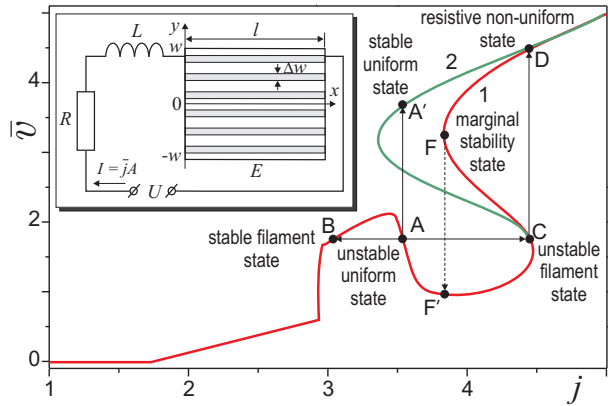}%[bb=94 490 434 704, width=8.0cm, clip]
\end{center}
\vspace{-1.5cm}
\caption{ (Color online) 
Schematic \textit{VI} curve of the superconductor
for two different values of the hysteresis due to overheating.
The more pronounced hysteresis loop (shown by the green solid line) corresponds
to a larger value of the characteristic heat release.
Inset: electrical circuit and the sample with filaments.}
\vspace{-0.5cm}
\end{figure}

For small values of $|\partial j/\partial B|$, the
solution of Eq.~(\ref{A9}) can be found explicitly
(with accuracy up to $|\beta^2|$): $\nu =-1-\gamma t_0/t_s$. It follows
from this relation that the instability occurs only at the
\textit{VI} curve branch with NDR, if the drop of the
voltage is large: $|\partial E/\partial j|>\rho_c$. The
characteristic size of the arising filamentary structure is of the
order of $\Delta w \propto w/|p_1|=w/\sqrt{|\gamma+t_s/t_0|}$. If
$| \gamma t_s/t_{0} | \gg 1$, the filament width is small, $\Delta w \ll
w$. Thus, the sample with NDR in the \textit{VI} curve of $S$ type
divides into small filaments with different current densities
(in different dynamic flux flow phases III and IV) at $Rl/A\ll
|\partial E/\partial j|$ and $L\ll 4\pi l/c^2A$. The obtained
results are valid if $\rho_c^{-1}|\partial E/\partial j|\gg [(4\pi
w/c)\partial j/\partial B]^2$. The instability occurs only if the
left hand side of this last inequality is higher than unity; the right
hand side is much smaller than 1 for the parameter range used
above if $w<1$~mm. In the stationary inhomogeneous state that
arises after development of the instability, the electric field
should be uniform over the sample.

A more complex dynamics appears when the \textit{VI} curve has
{\it simultaneously} both NDR parts of $N$ {\it and} $S$ types 
(see Fig.~2). In this case, the filaments with higher current density
$j_C$ are unstable if the system is far from the voltage-bias
regime~\cite{sch,GMR}. The instability of the filament with an
$N$ type \textit{VI} curve results in the switch of the filament
into the overheated state~\cite{GM} D, giving rise to a
non-uniform electric field distribution in the sample and, as a
result, to non-zero $\dot{B}$. Thus, the state that appears after
the instability develops is non-stationary. We consider two
possible \textit{VI} curves (type 1 red, type 2 green) shown in
Fig.~2. In state D the flux lines move fast, which is accompanied
with the acceleration of the flux flow in the lower-current
filaments. In the nearly current-biased mode~\cite{sch}, if the
\textit{VI} curve is of type 2, the sample comes to the uniform
stable state A$'$ with the current density $j_A$. However, if the
\textit{VI} curve has the form shown in the (red) curve 1, the
stable uniform state with the current density $j_A$ does not
exist. In this case the high electric field state moves from point
D to point F and falls down to a lower branch of the
\textit{VI} curve (point F$'$). In this state, the electric field
is also non-uniform and the lower and higher resistivity states
will move towards A. However, {\it this state is unstable and the
cycle of transitions will be repeated} ($A$ $\rightarrow$ $C$
$\rightarrow$ $D$ $\rightarrow$ $F$ $\rightarrow$ $F'$
$\rightarrow$ $A$). Such a cyclical dynamic state could be
realized in the form of flowing either stripes or resistive domain
walls moving along the filaments.

It is important to stress that the described cyclical 5-step
dynamics ($A$ $\rightarrow$ $C$ $\rightarrow$ $D$ $\rightarrow$
$F$ $\rightarrow$ $F'$ $\rightarrow$ $A$) obtained for the {\it
IV} curve having NDR of {\it both $N$ and $S$ types} cannot be
realized for {\it IV} curves with either {\it only N} or {\it
only S} type of NDR. The appearance of the non-stationary
oscillatory regime for stationary external conditions is very
unusual. This cyclic dynamics could be extended to different
physical systems, e.g., plasmas or superconductors without
artificial pins driven by a current flowing along the externally
applied magnetic field \cite{brandt}. Moreover, the predicted
dynamical behavior (which can be generalized for several other
systems, e.g., semiconductors) is potentially useful for the
transformation of a dc input into either an ac-current or a
voltage output which can be controlled by the parameters of the
external circuit. In a broader sense, this general class of cyclic
dynamics (ac output from dc input) is also found in other
important nonlinear systems, like the dc Josephson effect.

In summary,
for superconductors with periodic pinning arrays
with certain pinning and heat transfer characteristics,
we derive \textit{VI} curves with NDR of $N$ type, $S$ type, or {\it both}.
Complex dynamics and regimes
including domain structures and filamentary instabilities
appear when the \textit{VI} curve has both $S$ and $N$ types of NDR.
We analyzed the self-organized non-uniform dynamical regimes for
these instabilities.

This work was supported in part by ARDA and NSA under AFOSR
contract
F49620-02-1-0334; and by the US NSF grant No.~EIA-0130383.

\end{document}